\documentclass[a4paper]{article}

\usepackage[dvipsnames]{xcolor}  
\definecolor{titleblue}{rgb}{0.16,0.24,0.64} 
\definecolor{citecolor}{rgb}{0.2,0.3,0.8}

\usepackage{graphicx} 
\usepackage{mathtools}

\usepackage{hyperref}

\hypersetup{
    colorlinks    = true, 
    citecolor    = citecolor, 
    linkcolor    = red, 
    urlcolor    = magenta, 
}

\usepackage{breakurl}
\usepackage{stmaryrd}
\usepackage{wasysym}

\usepackage{multirow}

\makeatletter
\def\verbatim{\small\@verbatim \frenchspacing\@vobeyspaces \@xverbatim}
\makeatother

\makeatletter
\def\tt{}
\makeatother

\begin{document}


\title{Two-dimension tissue growth model based on circular granular cells for cells with small overlap}
\author{S. Viridi\thanks{viridi@cphys.fi.itb.ac.id}, S. N. Khotimah, D. Aprianti, L. Haris, and F. Haryanto \\[0.2cm]
Nuclear Physics and Biophysics Research Division \\
Physic Department, Faculty of Mathematics and Natural Sciences, \\
Institut Teknologi Bandung, Bandung 40132, Indonesia
}
\date{26 March 2014}
\maketitle

\begin{abstract}
Tissue growth can be modeled in two dimension by only using circular granular cells, which can grow and produce child. Linear spring-dashpot model is used to bind the cells with a cut-off interaction range of 1.1 times sum of radii of interacted cells. Simulation steps must be divided into explicit and implicit ones due to cell growing stage and cell position rearrangement. This division is aimed to avoid simulation problem. Only in the explicit steps time changes is performed. Large cells overlap is chosen as termination condition of tissue growth. Only some cells configuration can growth to infinite time without encountering the large cells overlap. These configurations, and the other also, are presented in this work. Simulation time $t$ increases as cell number $N$ increases due to raise of interaction number between two cells. Linear and network configurations tend to aligned with different asymtotic function, that relates $N$ and $t$, for large $N$.
\end{abstract}

\section{Introduction}

Model of tissue growth may be classified into phenomenological and mechanical types, where the first type attempts to simulate cause and effect without considering of the intermediary involved mechanical and biological mechanisms, while the later one begins with parameters that are linked to portions of the biological processes involved in tissue maintenance, turnover, and repair \cite{Cowin_2004}. Multiple approaches can used in a model of tissue growth, e.g. analytic using partial differential equation, cellular automaton, and equations such in a hybrid model, where each aproach is used for different purposes \cite{Cheng_2009}. In particular condition, additional equations must also included in the model, e.g. cell growth under direct perfusion, which requires also flow equations \cite{Chung_2007}. Normally, probability of a cell growth is provided by some probability function such in Monte-Carlo method \cite{Turner_2002}. Extending the complexity, mechanical properties such elasticity could also play also important role in constructing tissue forms such as leaves \cite{Dervaux_2008}. Even, environments for more complex system are already availabe, where one of the simulation environment for tissue growth, range from single-cell, multi-cell, organ, and organisms, is CompuCell3D \cite{Swat_2012}, which is also OpenSource with its predecessor is CompuCell \cite{Izaguirre_2004}. A very early simple model based only on circular cell \cite{Bodenstein_1986}, is adapted in this work, where near similar scheme can also be used to study simple motion of a cell \cite{viridi_2014}.

\section{Model}

At a particular time $t$ a cell $i$ is located at $\vec{r}_i(t)$. Supposed that it is represented by a circle with diameter $D_i(t)$, density $\rho$, and mass $m_i(t)$

\begin{equation}
\label{eqn:cg-cell-mass}
m_i(t) = \frac{1}{4}\rho \pi D_i^2(t),
\end{equation}

where growth of the cell is governed by

\begin{equation}
\label{eqn:cg-cell-diameter-growth}
D_i(t) = \left\{
\begin{array}{ll}
0, & t < t_{i0}, \\
\\
(t - t_{i0}) v_D, & {\displaystyle t_{i0} \le t \le t_{i0} + \frac{D_{\rm max}}{v_D}}, \\
\\
D_{\rm max}, & {\displaystyle t > t_{i0} + \frac{D_{\rm max}}{v_D}},
\end{array}
\right.
\end{equation}

with $t_{i0}$, $v_D$, and $D_{\rm max}$ stand for cell time of birth, cell diameter growth rate, and maximum cell diameter, respectively. It is assumed that all cells have same growth rate and maximum size. Age of a cell $\tau_i$ at time $t$ is can simply be found using

\begin{equation}
\label{eqn:cg-cell-age}
\tau_i = t - t_{i0},
\end{equation}

which should be positive. Negative value means that the cell is not yet born. Cells can start to reproce a child, when it is mature. Cell maturity $\mu$ is assumed related to maximum cell size. If maturity is define with value 0 (false) and 1 (true), then it can be represented using

\begin{equation}
\label{eqn:cg-cell-maturity}
\mu_i(t) = u\left(t_{i0} + \frac{D_{\rm max}}{v_D}\right),
\end{equation}

where $u$ is a step function,

\begin{equation}
\label{eqn:cg-step-function}
u(t) = \left\{
\begin{array}{ll}
0, & t < 0, \\
1, & \ge 0.
\end{array}
\right.
\end{equation}

If a cell $i$ can deliver more than one child, it needs a reproduction period $T$, which then triggers a birth of a new cell $j$

\begin{equation}
\label{eqn:cg-cell-birth-time}
t_{j0} = t_{i0} + \frac{D_{\rm max}}{v_D} + nT,~~n = 1, 2, ..,
\end{equation}

where $n$ indicates the $n$-th child. It is also assume that every cell has the same reproduction period. In Equation (\ref{eqn:cg-cell-birth-time}) $j$ is not necessary equal to $i + 1$, since $i$ and $j$ are indices of cells in the population. In a population with individu or cell that can only produce one child then $j = i + 1$, but not in other cases. Mother cell is labeled with $i$ and child cell is with $j$ in Equation (\ref{eqn:cg-cell-birth-time}). States, that indicates whether new cells can be reproduced or mother cell is fertil, can be also formulated using (\ref{eqn:cg-step-function}) and (\ref{eqn:cg-cell-birth-time})

\begin{equation}
\label{eqn:cg-cell-is-born}
\phi_j(t) = u(t_{j0}),
\end{equation}

which is similar to the representation of Equation (\ref{eqn:cg-cell-maturity}). There is also other parameter $C_{i, \rm max}$, which is number of children a cell $i$ allowed to reproduce. This value can be alternated zwischen mother cell and child cell. List of time related parameters for a cell is given in Table \ref{tab:cg-cell-time-parameters}. In this table $C_{i, \rm max}$ is not written explicit depent on time $t$, since it could be but it should not be.

\begin{table}
\center
\caption{Time related parameters of a cell.}
\label{tab:cg-cell-time-parameters}
\medskip
\begin{tabular}{cccc}
\hline
Symbol & Meaning & Uniqueness \\
\hline
$t_{i0}$ & birth time of cell $i$ & each cell \\
$\tau_i$ & age of cell $i$ & each cell \\
$T$ & reproduction period & none \\
$\mu_i(t)$ & cell maturity & each cell \\
$\phi_i(t)$ & cell fertility & each cell \\
$C_{i, \rm max}$ & allowed number of children & each cell \\
\hline
\end{tabular}
\end{table}

\bigskip

\begin{figure}[h]
\center
\includegraphics[width=5cm]{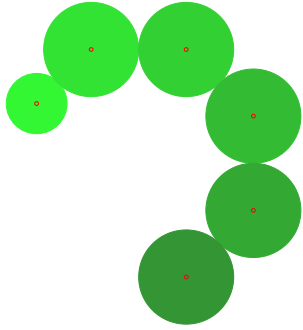}\caption{Configuration of cells, where each cell can only have one child and the child is always positioned at $\theta = \pi/4$ on mother cell circumference.}
\label{fig:example-th-pi4}
\end{figure}

As a cell, which refers to a mother cell, is mature and fertil, it can reproduce a child on some position along its circumference. Then Cell must also has an orientation $\theta_i$, which is used to position its new child relatively from, e.g. at $\theta_j$. Given in Figure \ref{fig:example-th-pi4} a configuration where child is alway put at $\theta = \pi/4$ from its mother orientation after the birth and every cell is only allowed to produce one child. Darker color means older cell. Child cell position on its mother circumference will be its orientation. It is a way how a child inherits one of its mother cell properties.

\bigskip
After its birth a child cell will occupy some space, it means that Equation (\ref{eqn:cg-cell-diameter-growth}) is lack of information how to position the child during its growth. A interaction force based on overlap between two cells must be formulated. The linear spring-dashpot, which is a common model for granular grains overlap \cite{Schaefer_1996}, can be adapted for this case as follow

\begin{equation}
\label{eqn:cg-cell-interaction-force}
\vec{F}_{ij}(t) = \left\{
\begin{array}{ll}
k \left\{ \displaystyle \frac12 [D_i(t) + D_j(t)] - r_{ij}(t) \right\} \hat{e}_{ij} - \gamma \vec{v}_{ij}(t), & r_{ij}(t) \le l_{ij}(t), \\
\\
0, & r_{ij}(t) > l_{ij}(t),
\end{array}
\right.
\end{equation}

for force act on cell $i$ because of its overlap with cell $j$, where $k$ and $\gamma$ are spring constant and dissipation factor, respectively. The first term in Equation (\ref{eqn:cg-cell-interaction-force}) for $r_{ij}(t) \le l_{ij}(t)$ represents binding force, which can be attractive or repulsive, while the second term is for dissipation to assure that the system does not ocsillate but goes to a static configuration. At time $t$ cell $i$ is at position $\vec{r}_i(t)$ with velocity $\vec{v}_i(t)$, while cell $j$ is at position $\vec{r}_j(t)$ with velocity $\vec{v}_j(t)$. Relative position and velocity of cell $i$ with respect to cell $j$ are

\begin{eqnarray}
\label{eqn:cg-cell-relative-position}
\vec{r}_{ij}(t) = \vec{r}_i(t) - \vec{r}_j(t), \\
\label{eqn:cg-cell-relative-velocity}
\vec{v}_{ij}(t) = \vec{v}_i(t) - \vec{v}_j(t).
\end{eqnarray}

\begin{figure}[h]
\center
\begin{tabular}{ccccc}
\includegraphics[width=2cm]{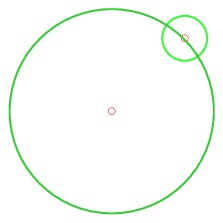} &
\includegraphics[width=2cm]{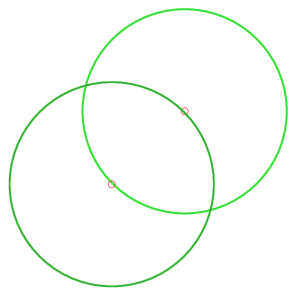} &
\includegraphics[width=2cm]{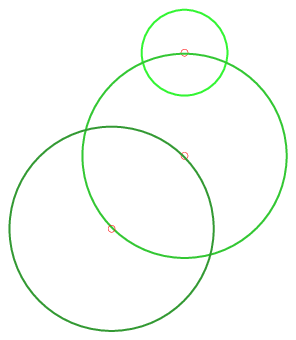} &
\includegraphics[width=2cm]{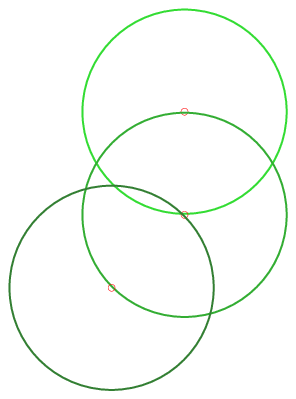} &
\includegraphics[width=2cm]{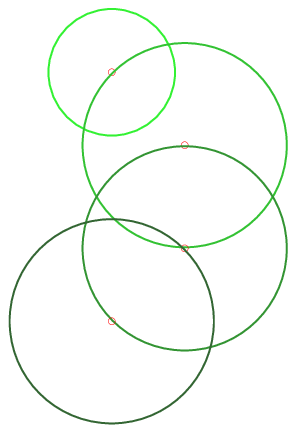} \\
(a) & (b) & (c) & (d) & (e) \\
\includegraphics[width=2cm]{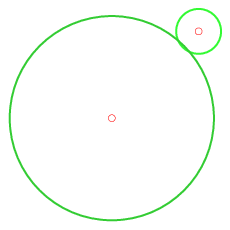} &
\includegraphics[width=2cm]{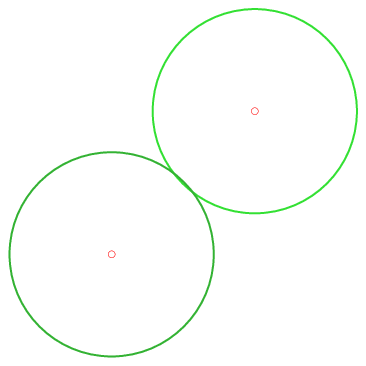} &
\includegraphics[width=2cm]{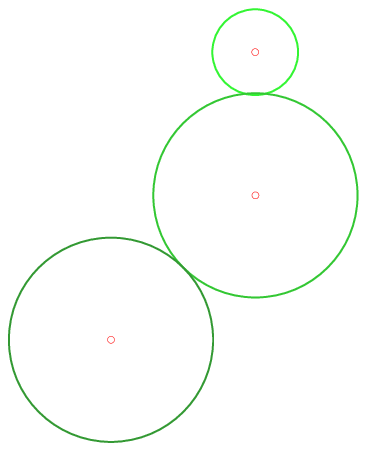} &
\includegraphics[width=2cm]{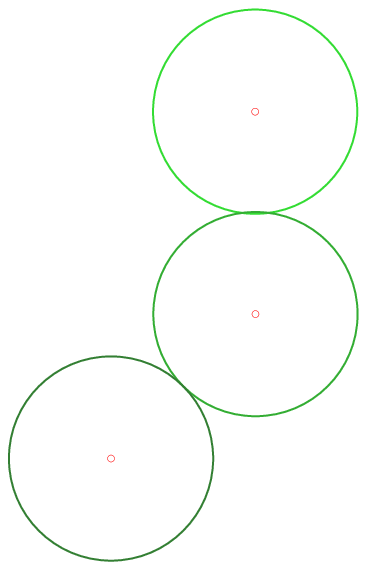} &
\includegraphics[width=2cm]{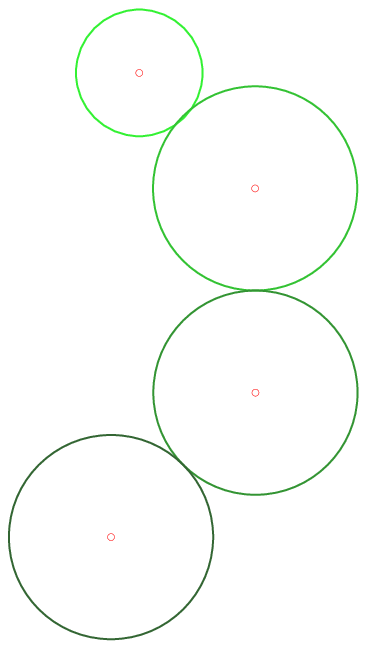} \\
(f) & (g) & (h) & (i) & (j) \\
\end{tabular}
\caption{Ilustration of cells position during growth: without (a-e) and with (f-j) interaction force $\vec{F}_{ij}$.}
\label{fig:force-0-1}
\end{figure}

In Equation (\ref{eqn:cg-cell-interaction-force}) there is a limit of interaction range $l_{ij}(t)$ between cell $i$ and cell $j$ to avoid long range interaction that could prevent the system to reach minimum energy. For only contact or very short range binding

\begin{equation}
\label{eqn:cg-cell-interaction-range}
l_{ij}(t) = \frac12 \alpha [D_i(t) + D_j(t)],
\end{equation}

with $\alpha$ near to 1. Influence of interaction force in Equation (\ref{eqn:cg-cell-interaction-force}) to the cells position is given in Figure \ref{fig:force-0-1}. Darker color indicates older cells, i.e. mother cell from cell with lighter color.

If total force suffered by cell $i$

\begin{equation}
\label{eqn:cg-cell-total-force}
\vec{F}_i(t) = \sum_{j \ne i} \vec{F}_{ij}(t),
\end{equation}

can be calculated, then acceleration of cell $i$ at time $t$ can be found using Newton second law of motion

\begin{equation}
\label{eqn:cg-cell-acceleration}
\vec{a}_i(t) = \frac{1}{m_i(t)} \vec{F}_i(t),
\end{equation}

where $m_i(t)$ is obtained from Equation (\ref{eqn:cg-cell-mass}). Among numeric integration methods, Euler method gives the simplest way to get new velocity and position at time $t + \Delta t$ for cell $i$ through

\begin{eqnarray}
\label{eqn:cg-cell-new-velocity}
\vec{v}_i(t + \Delta t) = \vec{v}_i(t) + \vec{a}_i(t) \Delta t, \\
\label{eqn:cg-cell-new-position}
\vec{r}_i(t + \Delta t) = \vec{r}_i(t) + \vec{v}_j(t) \Delta t.
\end{eqnarray}

New cells position obtained from Equation (\ref{eqn:cg-cell-new-position}) will fix unphysical results from Figure \ref{fig:force-0-1}.a-e to \ref{fig:force-0-1}.f-j, which more makes sense.

\bigskip

In the simulation scheme there are two major step: explicit and implicit steps. Explicit step is about cell growth and reproduction. The time changes in this step from $t$ to $t + \Delta t$. Between explicit steps there could be some implicit steps occur when some new cells are born since they occupy some spaces, which means that all cell must be rearranged. It is assumed that rearangement does not take time. Then in the implicit step the time $t$ remains the same. This assumption is based on the difficulties and time required to rearrange all cells. If time changes in this steps, which also induces new birth of other cells, then this step might not be finished. Competition between cell arrangement and cell birth is neglected in this work. And for the indicator of implicit step potential energy of the system must be calculated from each pair of cells $i$ and $j$

\begin{equation}
\label{eqn:cg-cell-pairs-potential-energy}
U_{ij}(t) = \left\{
\begin{array}{ll}
k \left\{ \displaystyle \frac12 [D_i(t) + D_j(t)]^2 - r_{ij}(t) \right\}^2, & r_{ij}(t) \le l_{ij}(t), \\
\\
0, & r_{ij}(t) > l_{ij}(t),
\end{array}
\right.
\end{equation}

using

\begin{equation}
\label{eqn:cg-cell-system-potential-energy}
U(t) = \frac12 \sum_{i \ne j} U_{ij}(t).
\end{equation}

Implicit steps will be terminated if $U(t) \le U_{\rm min}$.

\section{Results and discussion}

Following parameters are used in the simulation: $\Delta t = 10^{-3}$, $\rho = 1$, $T = 0.05$, $D_{\rm max} = 0.05$,  $v_D = 1$, $k = 100$, $\gamma = 1.8$, $U_{\rm min} = 10^{-5}$, $\alpha = 1.1$, $\theta = \pm \pi/3, \pm \pi/4,$ mixed, .. (see text), and $C_{i, \rm max}$ = 1, 2, mixed .. (see text).

\bigskip

\begin{figure}[h]
\center
\begin{tabular}{ccc}
\includegraphics[width=3.5cm]{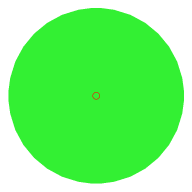} &
\includegraphics[width=3.5cm]{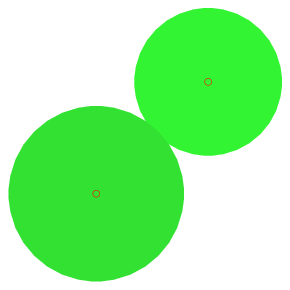} &
\includegraphics[width=3.5cm]{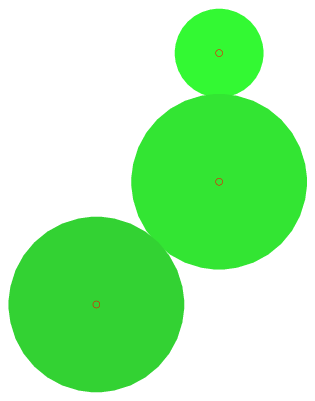} \\
$t$ = 0.058 & $t$ = 0.116 & $t$ = 0.174 \\
\includegraphics[width=3.5cm]{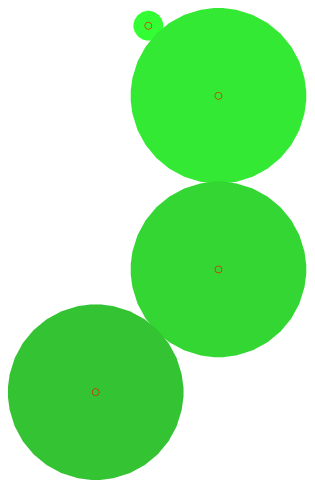} &
\includegraphics[width=3.5cm]{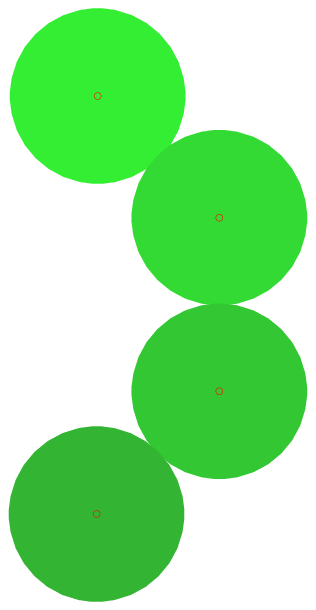} &
\includegraphics[width=3.5cm]{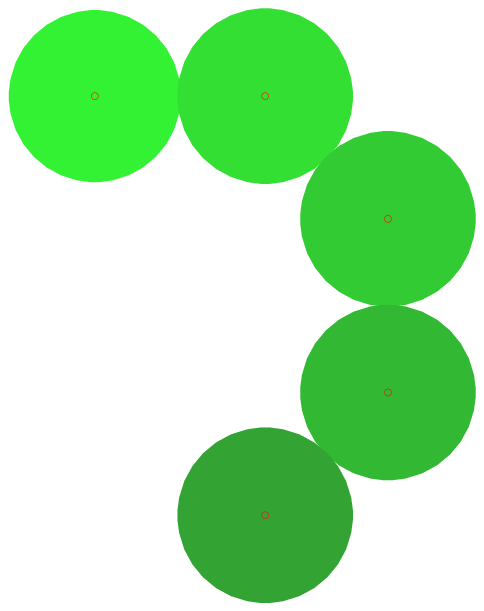} \\
$t$ = 0.232 & $t$ = 0.29 & $t$ = 0.348 \\
\includegraphics[width=3.5cm]{figures/cells-pi_div8-8-7} &
\includegraphics[width=3.5cm]{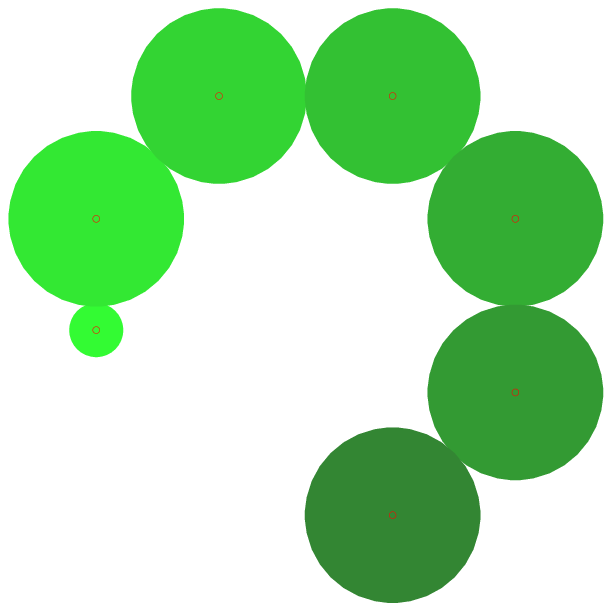} &
\includegraphics[width=3.5cm]{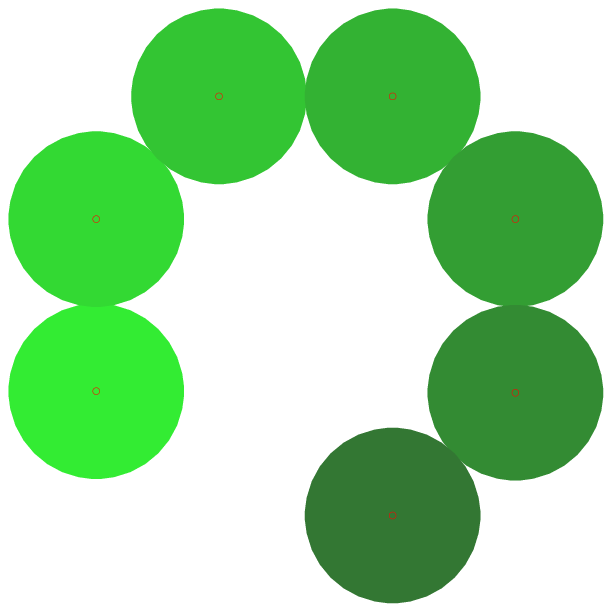} \\
$t$ = 0.406 & $t$ = 0.464 & $t$ = 0.522 \\
& \includegraphics[width=3.5cm]{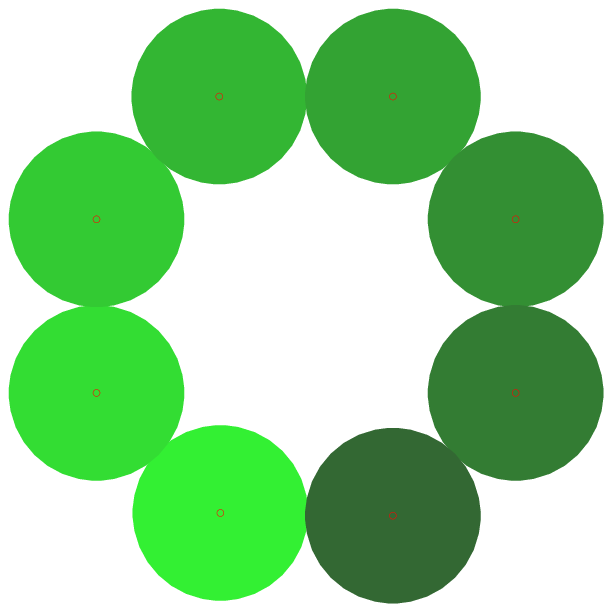} & \\
& $t$ = 0.58 & \\
\end{tabular}
\caption{An example of configuration consisted of eight cells that can only have one child with $\theta = \pi/4$, where older cell has darker color.}
\label{fig:ch1-pi4}
\end{figure}

\begin{figure}[h]
\center
\begin{tabular}{ccc}
\includegraphics[width=3.5cm]{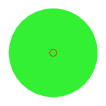} &
\includegraphics[width=3.5cm]{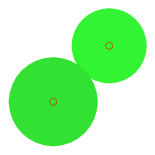} &
\includegraphics[width=3.5cm]{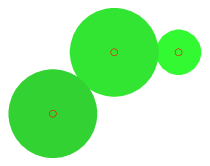} \\
$t$ = 0.058 & $t$ = 0.116 & $t$ = 0.174 \\
\includegraphics[width=3.5cm]{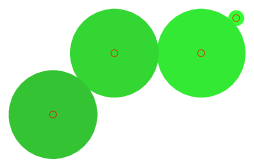} &
\includegraphics[width=3.5cm]{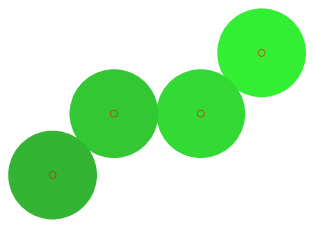} &
\includegraphics[width=3.5cm]{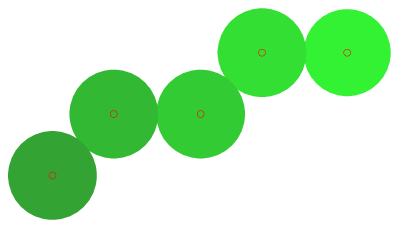} \\
$t$ = 0.232 & $t$ = 0.29 & $t$ = 0.348 \\
\includegraphics[width=3.5cm]{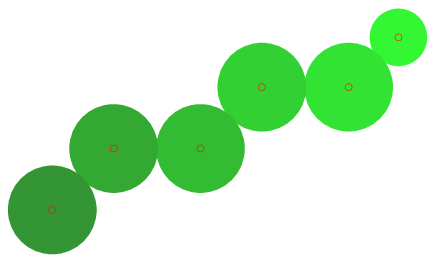} &
\includegraphics[width=3.5cm]{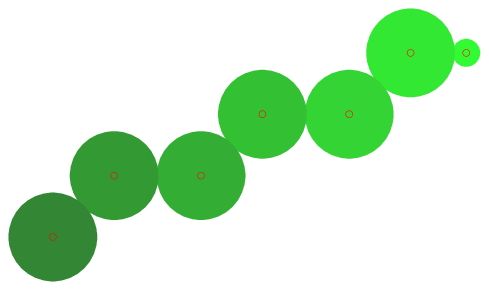} &
\includegraphics[width=3.5cm]{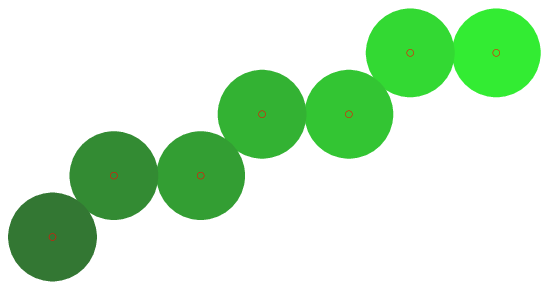} \\
$t$ = 0.406 & $t$ = 0.464 & $t$ = 0.522 \\
& \includegraphics[width=3.5cm]{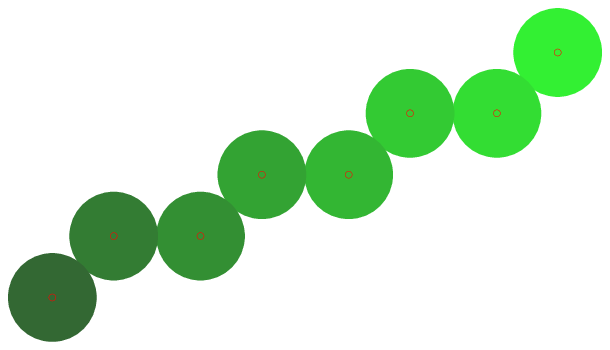} & \\
& $t$ = 0.58 & \\
\end{tabular}
\caption{An example of configuration consisted of eight cells that can only have one child with $\theta = \pm\pi/4$ (zig-zag), where older cell has darker color.}
\label{fig:ch1-pi4-zigzag}
\end{figure}

\begin{figure}[h]
\center
\begin{tabular}{ccc}
\includegraphics[width=3.5cm]{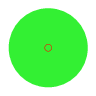} &
\includegraphics[width=3.5cm]{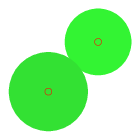} &
\includegraphics[width=3.5cm]{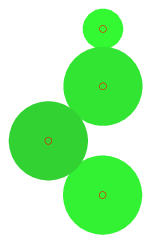} \\
$t$ = 0.058 & $t$ = 0.116 & $t$ = 0.174 \\
\includegraphics[width=3.5cm]{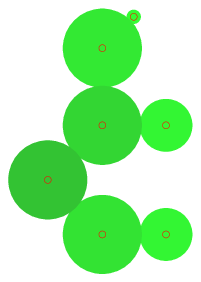} &
\includegraphics[width=3.5cm]{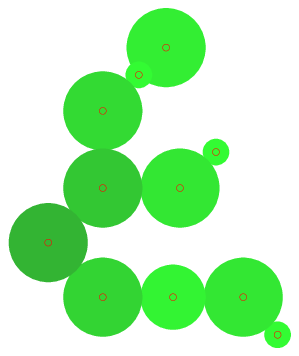} &
\includegraphics[width=3.5cm]{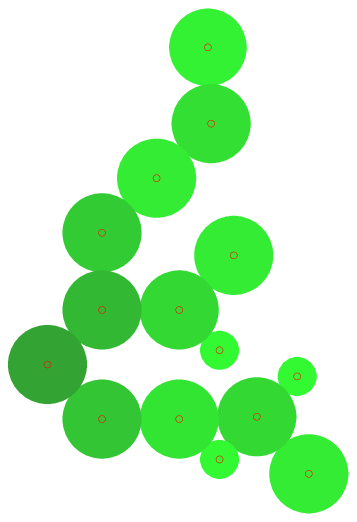} \\
$t$ = 0.232 & $t$ = 0.29 & $t$ = 0.348 \\
\end{tabular}
\caption{An example of configuration of cells that can only have two children with $\theta = \pm\pi/4$ (zig-zag), where older cell has darker color.}
\label{fig:ch2-pi4-zigzag}
\end{figure}

\begin{figure}[h]
\center
\begin{tabular}{ccc}
\includegraphics[width=3.5cm]{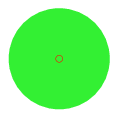} &
\includegraphics[width=3.5cm]{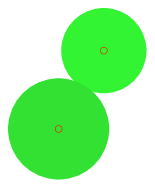} &
\includegraphics[width=3.5cm]{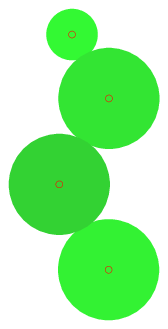} \\
$t$ = 0.058 & $t$ = 0.116 & $t$ = 0.174 \\
\includegraphics[width=3.5cm]{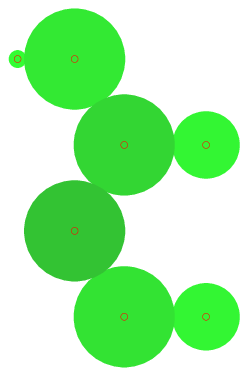} &
\includegraphics[width=3.5cm]{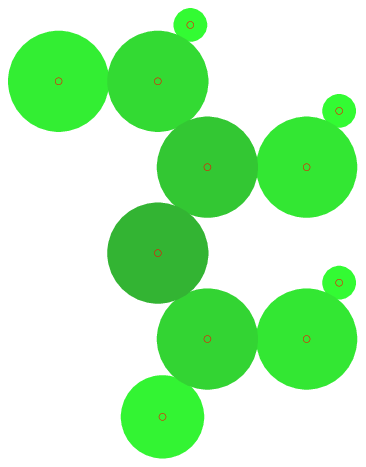} & \\
$t$ = 0.232 & $t$ = 0.29 &  \\
\end{tabular}
\caption{An example of configuration with cells that can only have two children with $\theta = \pm\pi/3$ for the first and second child, where older cell has darker color.}
\label{fig:ch2-pi3}
\end{figure}

\begin{figure}[h]
\center
\begin{tabular}{ccc}
\includegraphics[width=3.5cm]{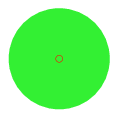} &
\includegraphics[width=3.5cm]{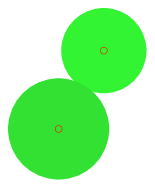} &
\includegraphics[width=3.5cm]{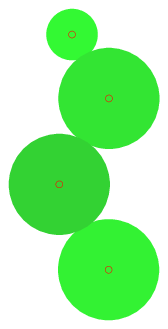} \\
$t$ = 0.058 & $t$ = 0.116 & $t$ = 0.174 \\
\includegraphics[width=3.5cm]{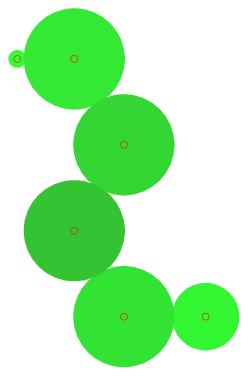} &
\includegraphics[width=3.5cm]{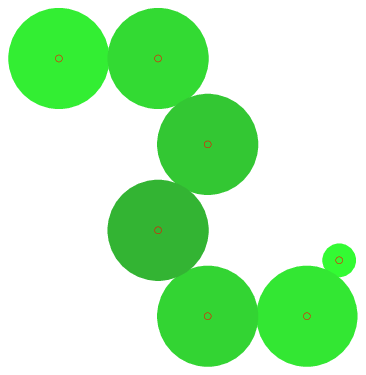} &
\includegraphics[width=3.5cm]{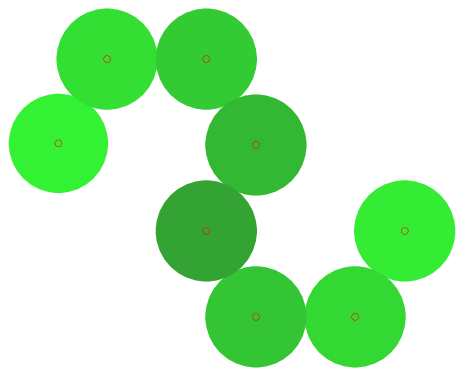} \\
$t$ = 0.232 & $t$ = 0.29 & $t$ = 0.348 \\
\includegraphics[width=3.5cm]{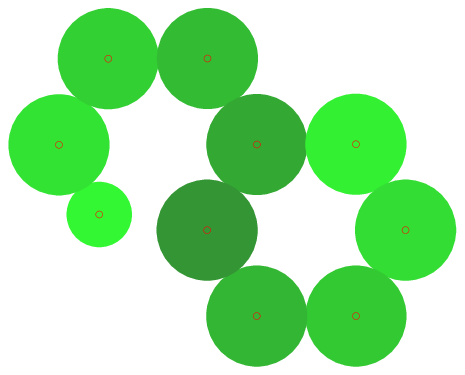} & & \\
$t$ = 0.406 & & \\
\end{tabular}
\caption{An example of configuration with first cell can have two children while the others can only have one with $\theta = \pi/3$ (first cell has $\theta = \pm\pi/6$), where older cell has darker color.}
\label{fig:ch1-ch2-pi3}
\end{figure}

\begin{figure}[h]
\center
\begin{tabular}{ccc}
\includegraphics[width=3.5cm]{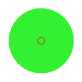} &
\includegraphics[width=3.5cm]{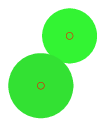} &
\includegraphics[width=3.5cm]{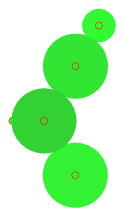} \\
$t$ = 0.058 & $t$ = 0.116 & $t$ = 0.174 \\
\includegraphics[width=3.5cm]{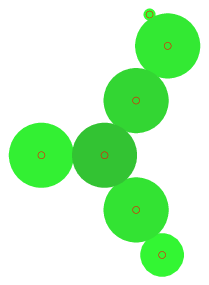} &
\includegraphics[width=3.5cm]{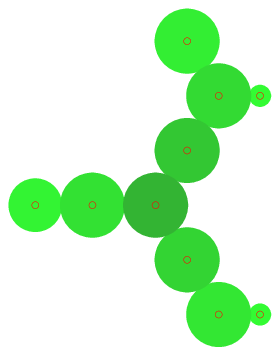} &
\includegraphics[width=3.5cm]{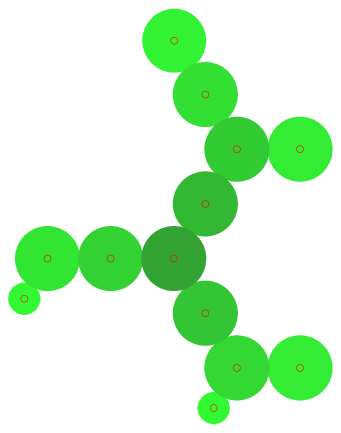} \\
$t$ = 0.232 & $t$ = 0.29 & $t$ = 0.348 \\
\includegraphics[width=3.5cm]{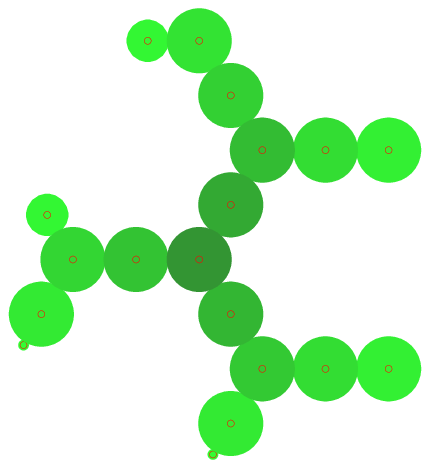} &
\includegraphics[width=3.5cm]{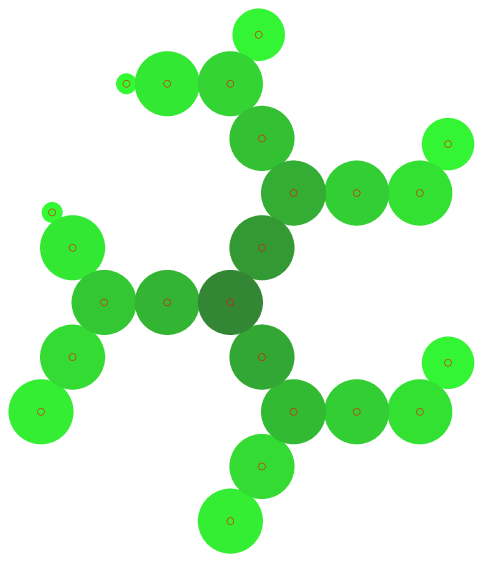} &
\includegraphics[width=3.5cm]{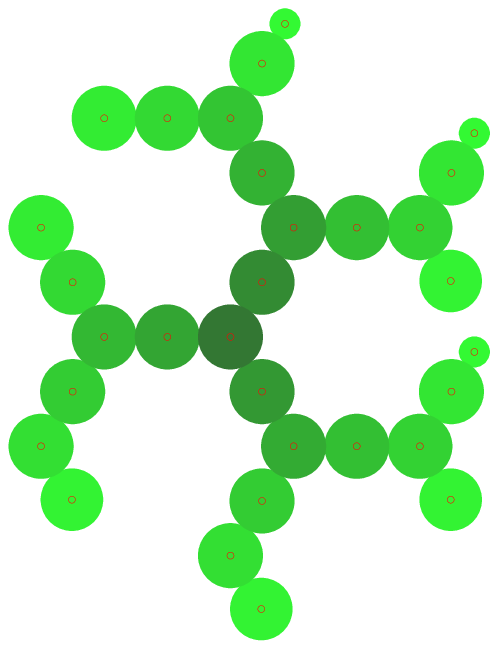} \\
$t$ = 0.406 & $t$ = 0.464 & $t$ = 0.522 \\
\end{tabular}
\caption{An example of configuration consisted cells that can only have three children, then one child, then two children, then one again .. (32121..) with $\theta = \pm\pi/6$ and $-\pi$ (the last only for the first cell), where older cell has darker color.}
\label{fig:ch31212-pi3}
\end{figure}

A rich configuration can be achieved by varying $\theta_i$ and $C_{i, \rm max}$ as the results are given in Figure \ref{fig:ch1-pi4} and \ref{fig:ch1-pi4-zigzag} only by change the constant $\theta_i$ into sqequence of $+\theta_i$, $-\theta_i$, $+\theta_i$, $-\theta_i$, .., where both configuration have $C_{i, \rm max} = 1$. Increasing number of children $C_{i, \rm max} = 2$ will produce branches as given in Figure \ref{fig:ch2-pi4-zigzag} and \ref{fig:ch2-pi3} for $\theta_i = \pi/4$ (zig-zag) and $\theta_i = \pi/3$ (for 1st and 2nd child), respectively. Other alternativ is if only the first cell can have two children but the other can only have one, with $\theta_i = \pm\pi/3$ for first cell and $\theta_i = \pi/3$ for other cells, the result is given in Figure \ref{fig:ch1-ch2-pi3}. More complex sytem can also be developed, e.g. sequence of number of children $C_{i, \rm max}$ = 3, 2, 1, 2, 1, .. and sequence $\theta_i = (\pm\pi/3, \pi), \pi/3, \pi/3, ..$, such shown in Figure \ref{fig:ch31212-pi3}.

\bigskip

\begin{figure}[h]
\center
\begin{tabular}{cc}
\includegraphics[width=5cm]{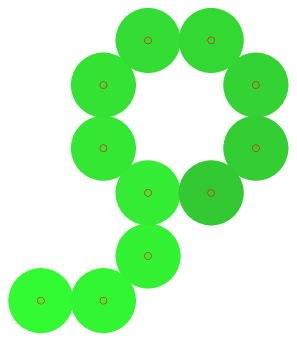} &
\includegraphics[width=5cm]{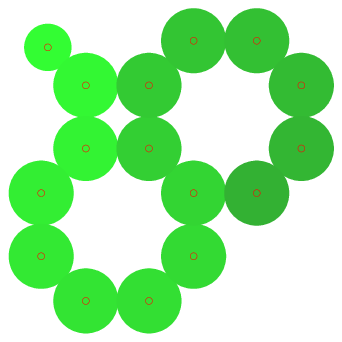} \\
$t$ = 0.812 & $t$ = 1.16  \\
\includegraphics[width=5cm]{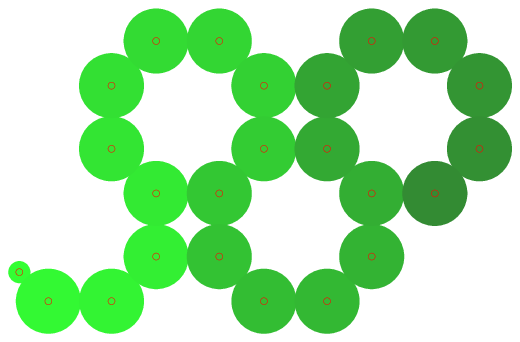} &
\includegraphics[width=5cm]{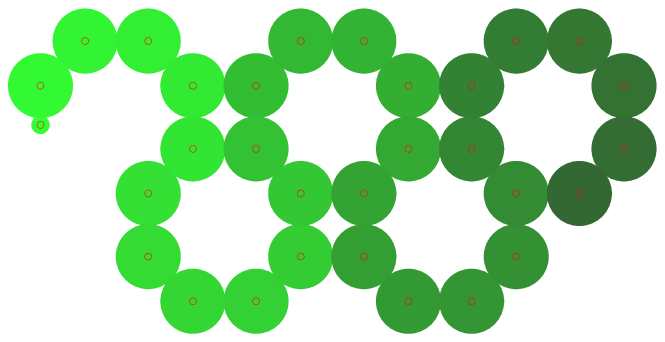} \\
$t$ = 1.74 & $t$ = 2.262 \\
\end{tabular}
\caption{An example of configuration consisted cells that can only have one child with $\theta = \pi/6$ for first 8 cells, $-\pi$ for next 6 cells, $\pi$ for next 6 cells, .., where older cell has darker color.}
\label{fig:ch1-pi4-7p-6n-6p-6n}
\end{figure}

Recent work has not yet considered large overlap, which happens when a child comes up between two existing cells. This could lead to break of the system since new position arrangement provided by Equations (\ref{eqn:cg-cell-total-force}) - (\ref{eqn:cg-cell-new-position}) can not accomodate this. Small $\Delta t$ is needed but it will be too luxurious for the explicit steps. Configuration in Figure \ref{fig:ch1-pi4}, \ref{fig:ch2-pi4-zigzag}, \ref{fig:ch2-pi3}, \ref{fig:ch1-ch2-pi3}, and \ref{fig:ch31212-pi3} will be terminated due to this large overlap. Only configuration in Figure \ref{fig:ch1-pi4-zigzag} and \ref{fig:ch1-pi4-7p-6n-6p-6n} can survive to grow until infinite time.

\bigskip

\begin{figure}[h]
\center

\includegraphics[width=11cm]{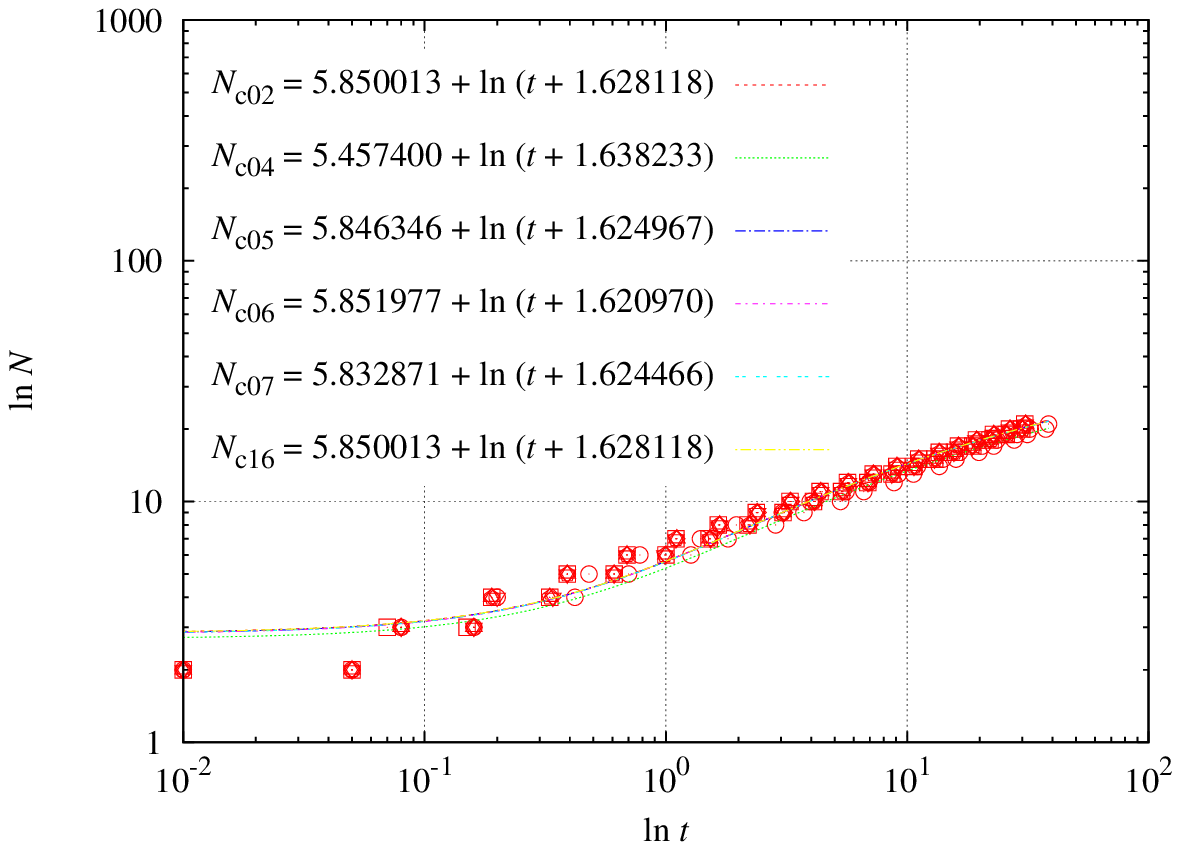}
\caption{Asymtotic feature of $N$ as function of $t$ for  linear configurations (refer Table \ref{tab:others} for the meaning of {\tt cxy} code).}
\label{fig:cxx-linear}

\bigskip

\includegraphics[width=11cm]{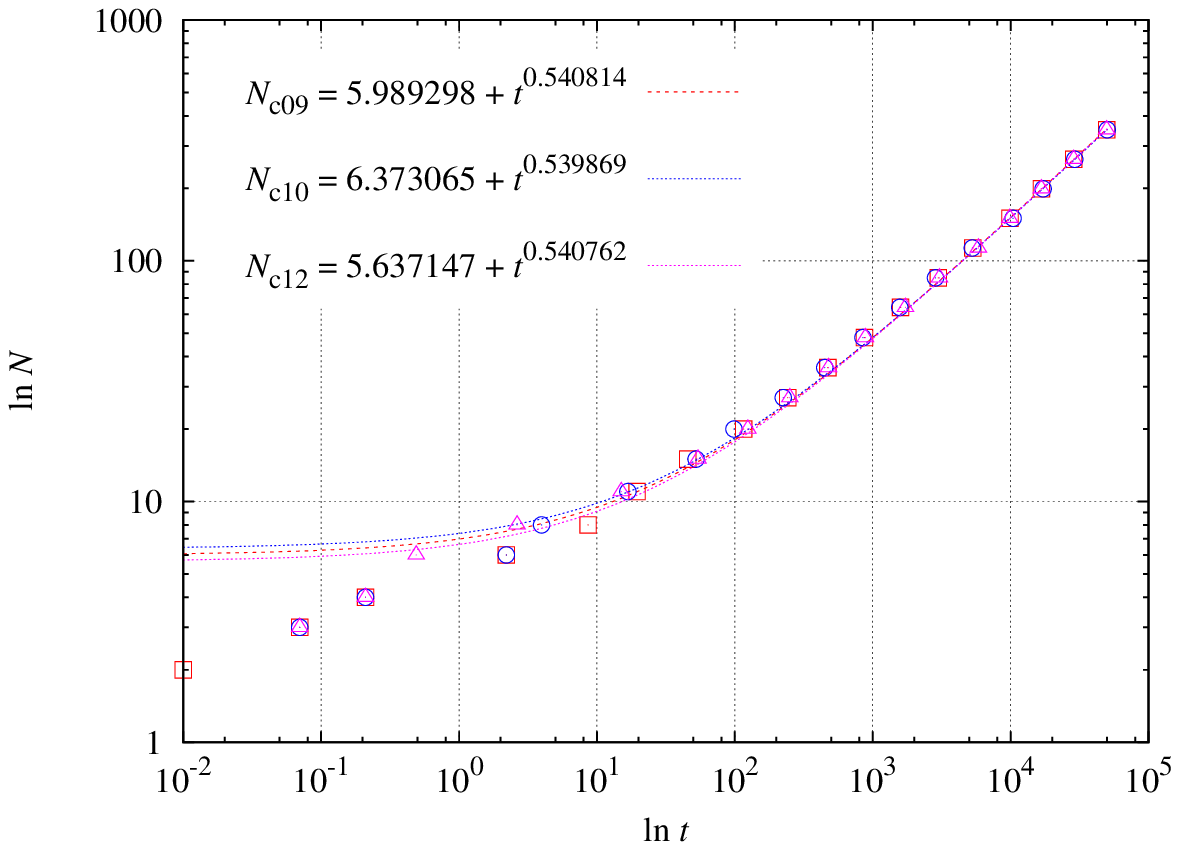}
\caption{Asymtotic feature of $N$ as function of $t$ for network configurations (refer Table \ref{tab:others} for the meaning of {\tt cxy} code).}
\label{fig:cxx-network}
\end{figure}

It is also observed that more overlaps occures between cells more real simulation time required, e.g for linear configurations the time required are almost the same as shown in Figure \ref{fig:cxx-linear} , but they will be different for network configurations as shown in Figure \ref{fig:cxx-network}. Both types of configuration produce asymtotic functions, which are

\begin{eqnarray}
\label{eqn:asymtote-linear}
N_{\rm linear}(t) = a + \ln (t + b) \\
\label{eqn:asymtote-network}
N_{\rm network}(t) = a + t^b.
\end{eqnarray}

Unfortunately, both functions are failed for small value of $N$.

\bigskip

Next plan is how to overcome the problem with large overlap between cells by using better method for cells rerangement instead of Euler method.It is also possible to add new rule that child can not be produced or can not grow if large overlap exists. More interesting results could come up since the tissue could change due to this large overlap.

\section{Conclusion}

A model for two dimension tissue growing based on circular granular cell has been presented, which can show rich configuration as long as the cells do not have large overlap with each other. From first nine calculated configurations only two can produce large tissue without introducing large overlap between cells, where these configuration can be considered as linear networks. Two asymtotic fucntions have been proposed for relating $N$ and $t$, each for linear and network configurations.

\bibliographystyle{unsrt}
\bibliography{refs}

\newpage
\section*{Appendix 1}

\subsection*{Parameters file}

For the code {\tt cellsgrow3} a parameters file is required, which normally named as {\tt params.txt}

\begin{verbatim}
# Iteration parameters
TMIN	0.0
TMAX	4.0
DT	1E-2

# Cell parameters
RHO	1.0
DMAX	0.05
VD	1.0
KR	1E2
KV	1.8
UMIN	1E-5
IUMAX	1E5

# Reproduction parameters
TPERIOD	0.05
MAXCHILD	1
THETA	0.1	0.2	0.1	0	-0.1	-0.2	-0.1	0

# Prefix for output filename and its producing period
TSHOW	0.05
PREFIX	output-
\end{verbatim}

\begin{figure}[h]
\center
\begin{tabular}{cccc}
\includegraphics[width=3cm]{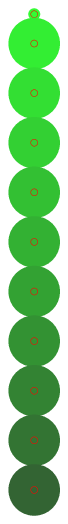} &
\includegraphics[width=3cm]{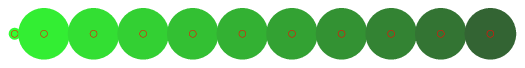} &
\includegraphics[width=3cm]{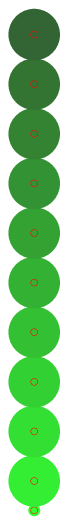} &
\includegraphics[width=3cm]{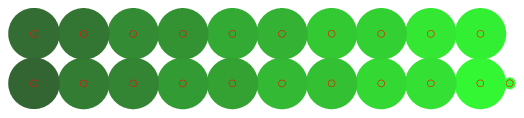} \\
c01 & c02 & c03 & c04 \\
\includegraphics[width=3cm]{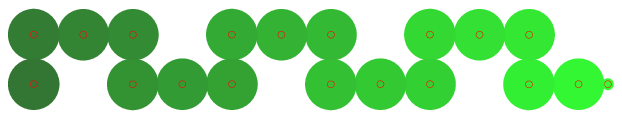} &
\includegraphics[width=3cm]{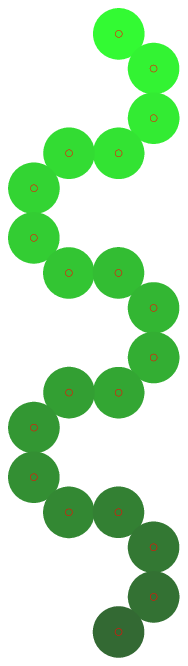} &
\includegraphics[width=3cm]{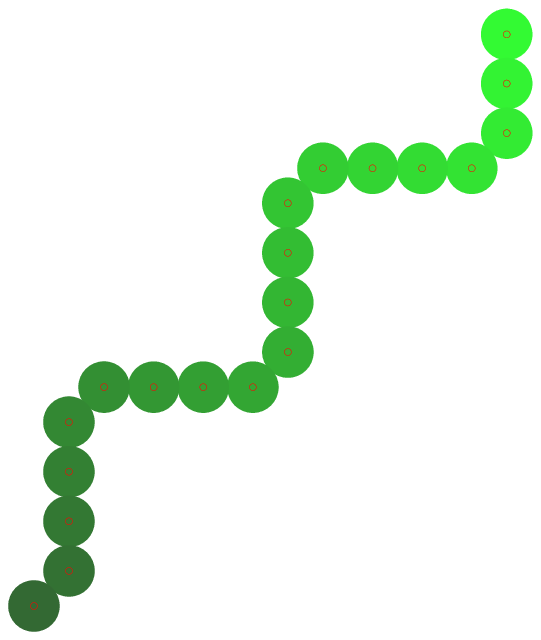} &
\includegraphics[width=3cm]{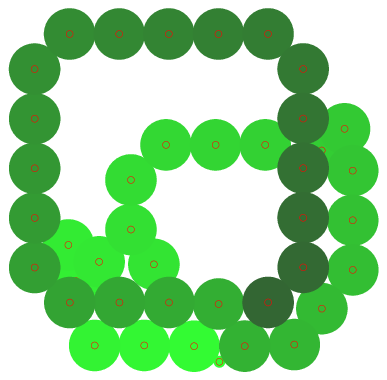} \\
c05 & c06 & c07 & c08 \\
\includegraphics[width=3cm]{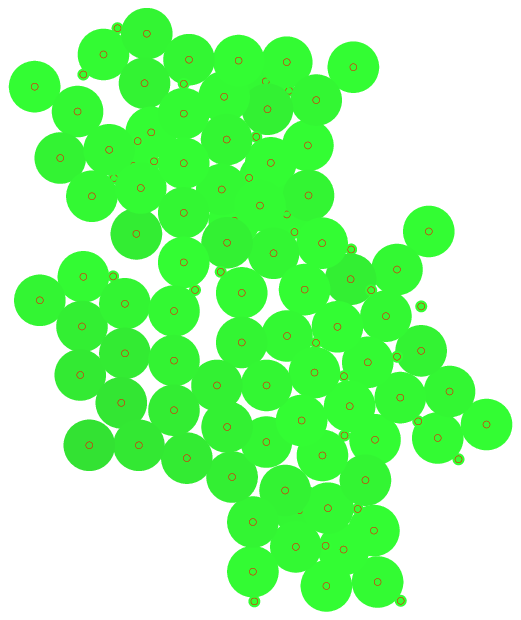} &
\includegraphics[width=3cm]{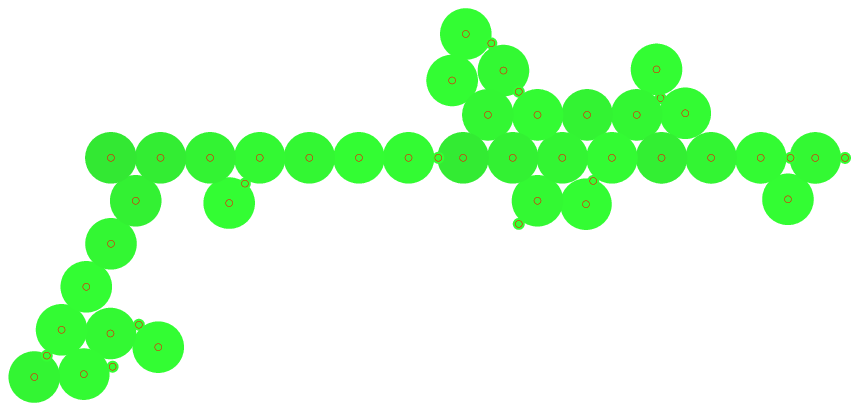} &
\includegraphics[width=3cm]{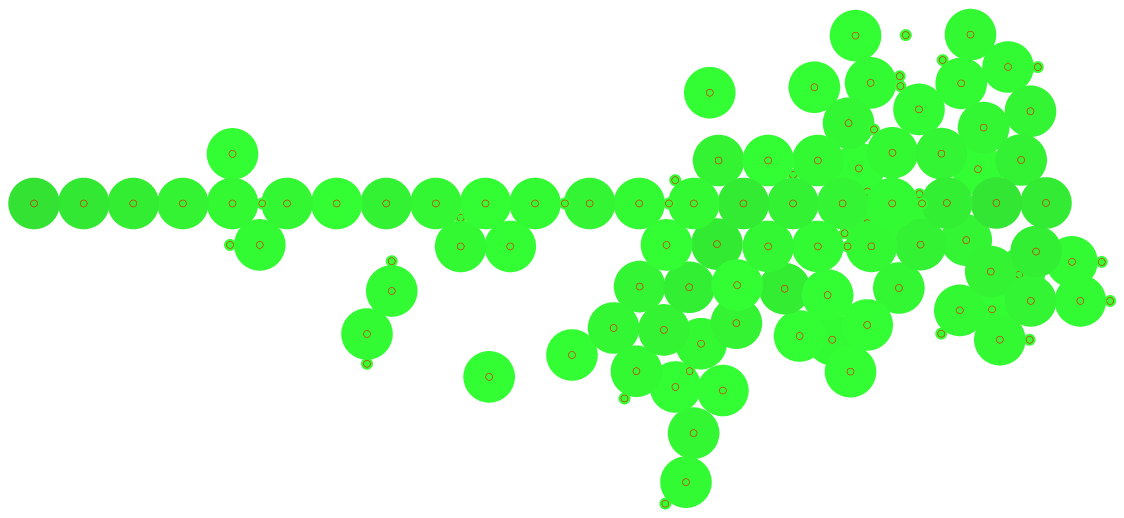} &
\includegraphics[width=3cm]{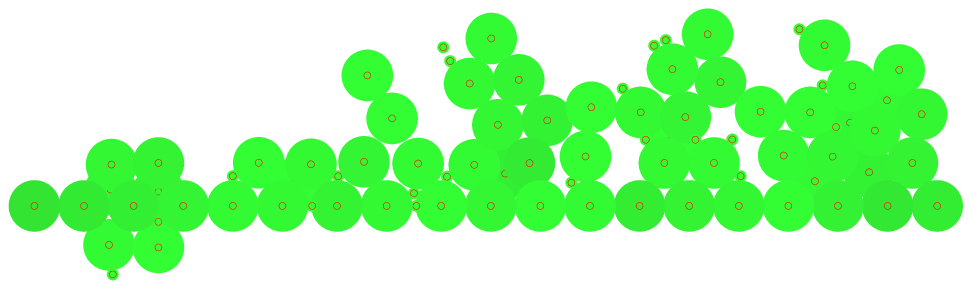} \\
c09 & c10 & c11 & c12 \\
\includegraphics[width=3cm]{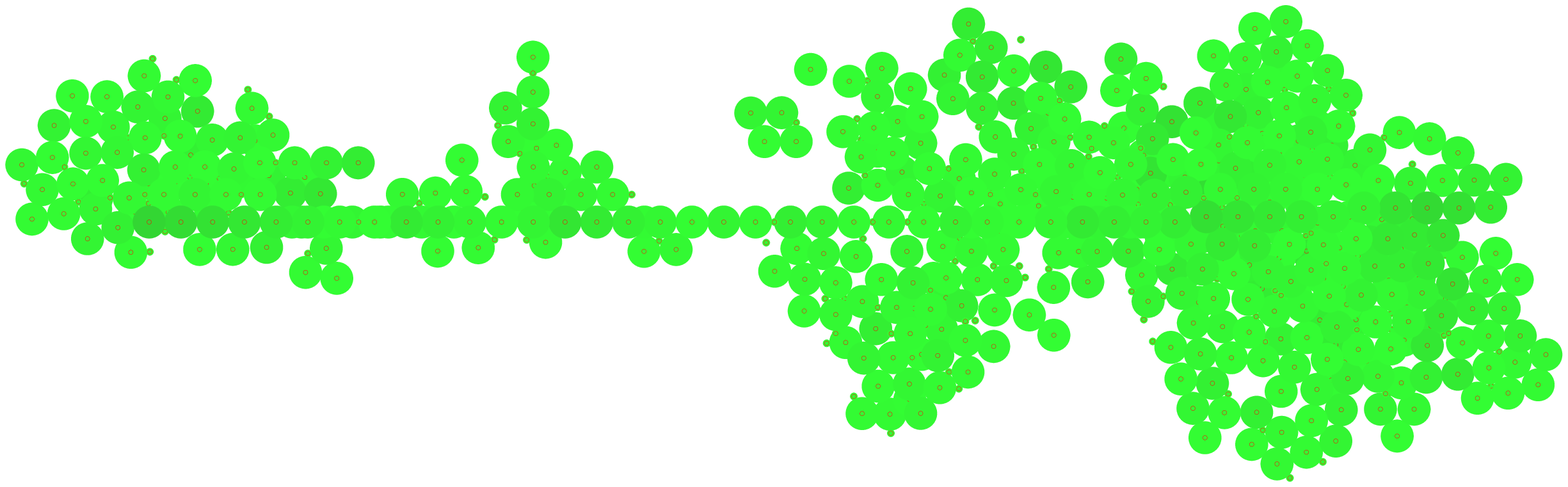} &
\includegraphics[width=3cm]{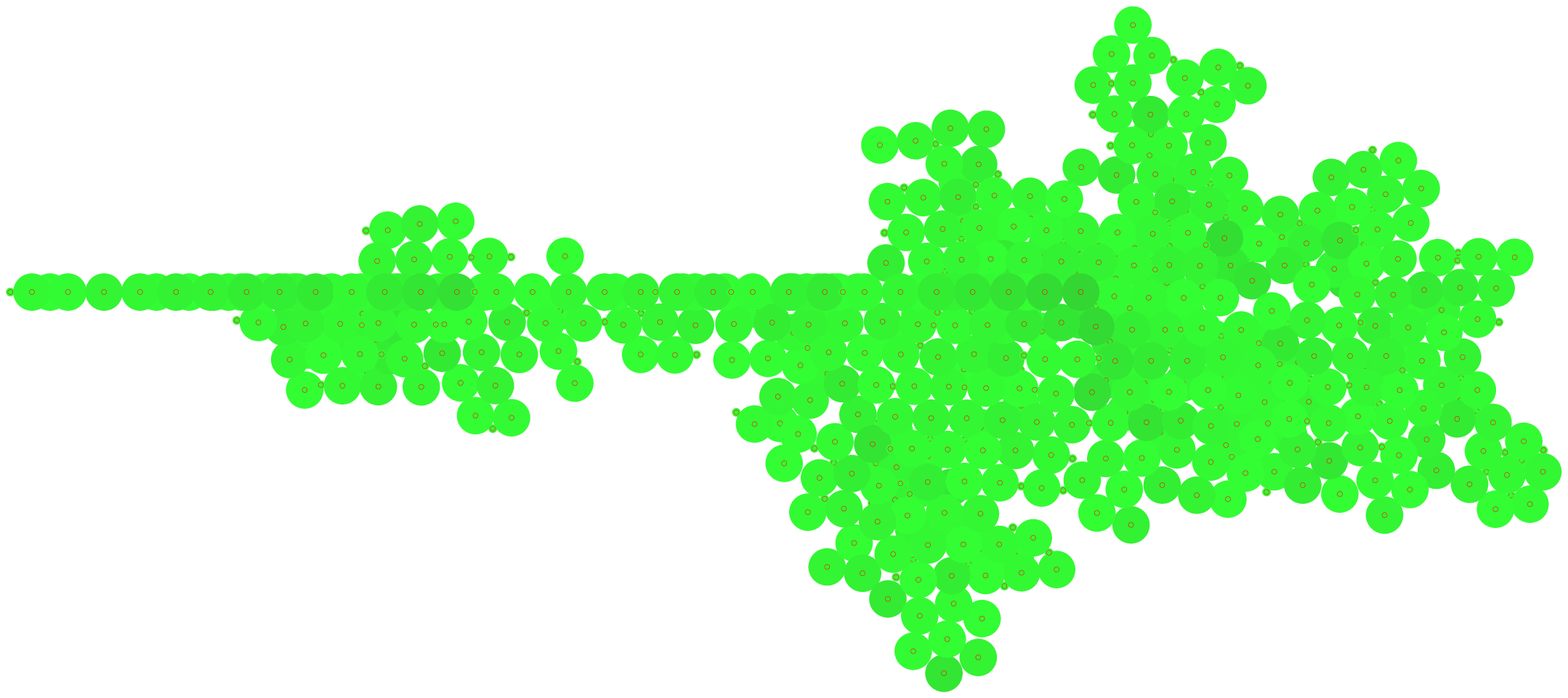} &
\includegraphics[width=3cm]{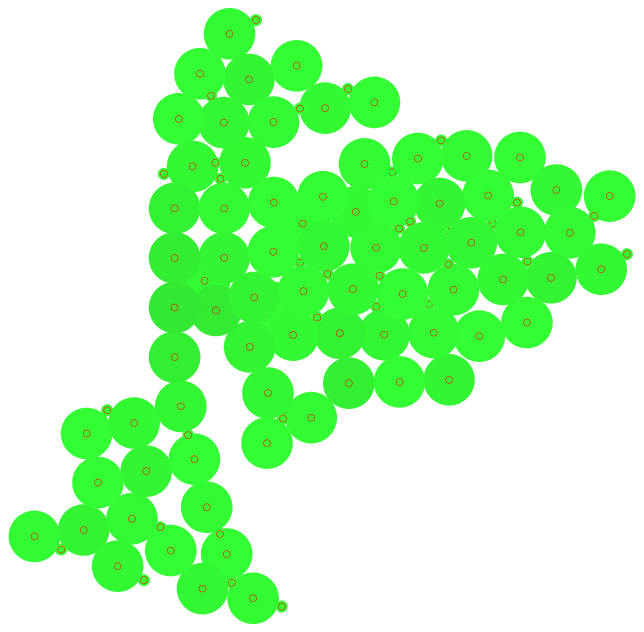} &
\includegraphics[width=3cm]{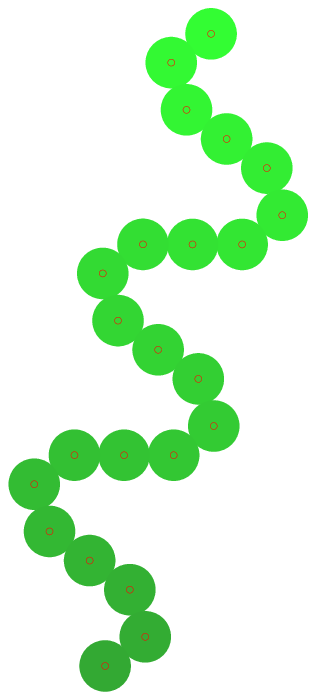} \\
c13 & c14 & c15 & c16 \\
\end{tabular}
\caption{Other examples of configurations with sequence defined in {\tt params.txt} as listed in Table \ref{tab:others}.}
\label{fig:others}
\end{figure}

Examples of sequence, which is given on {\tt THETA} in parameters file are shown in Figure \ref{fig:others} and Table \ref{tab:others}.

\begin{table}
\center
\caption{Parameters for Figure \ref{fig:others}.}
\label{tab:others}
\medskip
\begin{tabular}{ccccc}
\hline
Configuration & Children / cell & $\Delta t$ & $t_{\max}$ & Sequence of $\theta$ \\
\hline
c00 & 1 & 0.01 & 1 & 0 \\
c01 & 1 & 0.01 & 1 & 0.25	0	0	0	0	0	0	0	0	0 \\
c02 & 1 & 0.01 & 1 & 0.5	0	0	0	0	0	0	0	0	0 \\
c03 & 1 & 0.01 & 1 & 0.75	0	0	0	0	0	0	0	0	0 \\
c03 & 1 & 0.01 & 1 & 0.25	-0.25 \\
c04 & 1 & 0.01 & 2 & 0.25	-0.25	-0.25	0.25 \\
c05 & 1 & 0.01 & 2 & 0.25	-0.25	0	-0.25	0.25	0 \\
c06 & 1 & 0.01 & 2 & 0.125	0.125	0.125	0.125 \\
& & & & -0.125	-0.125	-0.125	-0.125 \\
c07 & 1 & 0.01 & 2 & 0.125	0.125	0	0	-0.125	-0.125	0	0 \\
c08 & 1 & 0.01 & 1 & 0.125	0.125	0	0	0 \\
c09 & 2 & 0.01 & 0.75 & 0.125	0	0	-0.125	0	0 \\
c10 & 2 & 0.01 & 0.6 & 0	0	0	0	-0.333	0.333 \\
c11 & 2 & 0.01 & 0.75 & 0	0	-0.25	0	0	0.25 \\
c12 & 2 & 0.01 & 0.7 & 0	0	0.25 \\
c13 & 2 & 0.01 & 1.05 & 0	0	0.5	0	0	-0.25	0	0	0.5 \\
c14 & 2 & 0.01 & 1.05 & 0	0.5	0	-0.5 \\
c15 & 2 & 0.01 & 0.6 & 0.02	0.25	0.02	-0.25	0.02 \\
c16 & 1 & 0.01 & 2.25 & 0.1	0.2	0.1	0	-0.1	-0.2	-0.1	0 \\
\hline
\end{tabular}
\end{table}

\end{document}